\documentclass[a4paper,fleqn,usenatbib]{mnras}

\usepackage{newtxtext,newtxmath}

\usepackage[T1]{fontenc}
\usepackage{ae,aecompl}

\usepackage{graphicx}	
\usepackage{amsmath}	
\usepackage{amssymb}	

\title[]{Three-dimensional simulations of accretion flow in the progenitor of Tycho's supernova}

\author[L. Xue et al.]{
Li Xue$^{1}$\thanks{E-mail: lixue@xmu.edu.cn},
Cheng-Liang Jiao$^{2}$\thanks{E-mail: jiaocl@ynao.ac.cn},
and Yuan Li$^{1}$
\\
$^{1}$Department of Astronomy, Xiamen University, Xiamen, Fujian 361005, P. R. China\\
$^{2}$Yunnan Observatories, Chinese Academy of Sciences, Kunming, Yunnan 650216, P. R. China\\
}

\date{Accepted XXX. Received YYY; in original form ZZZ}

\pubyear{2019}

\begin{document}
\label{firstpage}
\pagerange{\pageref{firstpage}--\pageref{lastpage}}
\maketitle

\begin{abstract}
We run three-dimensional numerical simulations for the accretion flow around the white dwarf (WD) in the progenitor system of Tycho's supernova (SN). The mass of the WD, mass of the companion star, and the orbital period are set to be 1$M_{\odot}$, 1.6$M_{\odot}$, and 0.794 day, respectively, based on theoretical and observational researches of Tycho's SN remnant (SNR). We find that when the magnetic field in the accreted material is negligible, outflowing wind is concentrated near the equatorial plane. When the magnetic field has energy equipartition with internal energy, polar wind is comparable with the equatorial wind. A carefully chosen magnetic field between the above two cases ($B=5.44\times10^3 \rm{G}$) can roughly reproduce the latitude-dependent wind required to form the peculiar periphery of Tycho's SNR.
\end{abstract}




\section{Introduction}

It is now widely accepted that Type Ia supernovae (SNe Ia) are thermonuclear explosions of mass-accreting carbon-oxygen white dwarfs (CO WDs) which have grown to the Chandrasekhar mass limit \citep[e.g.,][]{Hoyle1960, Nomoto1984, Nomoto1997}. Theoretical studies indicate that SNe Ia can be triggered by a CO WD accreting sufficient material from a non-degenerate companion star \citep[single-degenerate model, hereafter SD model;][] {Whelan1973, Nomoto1984, Fedorova2004, Han2008}. There are also other progenitor models for SNe Ia, such as the double-degenerate model \citep[e.g.,][]{Webbink1984, Iben1984}, the sub-Chandrasekhar mass model \citep[e.g.,][]{Nomoto1982, Woosley1986, Livne1990, Branch1995, Hoeflich1996, Sim2010}, etc., and the collection of observed SNe Ia events may be results from different origins \citep[see the review by][]{Wang2018}. Nevertheless, the SD model is supported by recent observations \citep[e.g.,][]{Patat2007, Wang2009, Sternberg2011, Dilday2012, Silverman2013a, Silverman2013b, Cao2015, Liu2015, Marion2016} and plays an important role in the production of SNe Ia.

The historical Tycho's SN (SN 1572) is thought to be Type Ia as evidenced by its light curve \citep{Baade1945, Ruiz-Lapuente2004} and X-ray spectroscopic analysis of its remnant \citep{Badenes2006, Krause2008}. For its progenitor system, the SD model is favored by a series of works \citep{Ruiz-Lapuente2004, Han2008, Wang+Han2010, Lu2011, Zhou2016, Fang2018} though it is still being debated \citep[e.g.,][]{Badenes2007, Krause2008, Woods2017}. \cite{Ruiz-Lapuente2004a} discoveried a fast-moving ($\sim 136\rm{km}/s$), type G0-G2 subgiant (Tycho-G) in Tycho's SNR and suggested it as the surviving companion of the SN, which likely was a main-sequence star or a slightly evolved subgiant in the progenitor of Tycho's SN. Other candidates of the surviving companion are also proposed, such as Tycho-B \citep{Kerzendorf2013} and Tycho-E \citep{Ihara2007}. However, Tycho-G is supported by \cite{Han2008} \citep[see also][]{Wang+Han2010}, who studied the evolution of a binary composed of a WD and a main-sequence star (WD+MS channel) until SN Ia explosion and gave various properties of the surviving companion, which are compatible with the observed properties of Tycho-G. \cite{Lu2011} claimed that the non-thermal X-ray arc in Tycho's SNR may originate from the interaction between SN ejecta and an envelope formed by stripped mass from the companion. \cite{Zhou2016} suggested that the expanding bubble surrounding Tycho's SNR is driven by a fast outflow from the vicinity of the WD as it accretes matter from its non-degenerate companion based on observations. Recently, \cite{Fang2018} argued that it provides an explanation for the peculiar shape of the periphery of Tycho's SNR that the SN ejecta evolved in a bubble blown by a latitude-dependent wind, of which the configuration was originally proposed by \cite{Rest2008} and was also employed to simulate the wind-driven bubble around the Kepler SNR \citep{Toledo-Roy2014} and the Cygnus loop \citep{Fang2017}. 

The above mentioned studies provided a convincing explanation for the progenitor system of Tycho's SN, in which a WD accretes matter from a non-degenerate companion (likely a main-sequence star) and blows out a bubble with its anisotropic outflow until the SN Ia explosion, and the shape of the SNR is formed by the SN ejecta evolving in the bubble. But is the special configuration of the anisotropic outflow used by \cite{Fang2018} natural or even possible? And if it is possible, can we constrain the properties of the SN progenitor using this special configuration? To answer these questions, we perform simulations of the accretion process within the WD's Roche lobe with a modified version of the code Athena 4.2 \citep{Stone2008}. The sturcture of the anisotropic outflow is obtained by simulation and compared with the configuration used by \cite{Fang2018}, and the properties of the SN progenitor system which can form such an outflow are discussed. Our work revisits the model of \cite{Makita2000}, but some adjustments have been made to fit the coding environment of Athena 4.2, and we focus on different range of parameters than \cite{Makita2000}. Our work is also different from \cite{Ju2016,Ju2017}, which are in cylindrical coordinates and performed with Athena++, a complete re-write of the Athena code in C++. In either \cite{Makita2000} or \cite{Ju2016,Ju2017}, the model only considered the region close to the equatorial plane and the vertical range of the computational domain is very limited, because their studies are focused on a supposedly thin accretion disc. In contrast, our model contains the entire Roche lobe of the WD and can be used to investigate the outflow structure.

The paper is arranged as follows. In Section \ref{Sec:Model}, we introduce the basic equations and input parameters used in our simulations. In Section \ref{Sec:Method}, we discuss the numerical method in our calculation, including the grid setting, the initial and boundary conditions and modifications we made to Athena 4.2 tailored to our needs. The results of five different cases of simulation are presented in Section \ref{Sec:Result}. We summarize and discuss our results in Section \ref{Sec:Summary}.

\section{The Model}\label{Sec:Model}
\subsection{Basic equations}
We revisit the three-dimensional (3D) model of \cite{Makita2000} and follow them to use a Cartesian frame corotating with the binary system. We only consider the inviscid flow, in which the gas loses its angular momentum with spiral shock, an essential mechanism of angular momentum transport confirmed by observational and numerical studies \citep[e.g.,][]{Kononov2012, Ju2016}. The basic equations are described in vectorized form as
\begin{equation}
\frac{\partial \bar{Q}}{\partial t}+\frac{\partial \bar{E}}{\partial x}+\frac{\partial \bar{F}}{\partial y}+\frac{\partial \bar{G}}{\partial z}+\bar{H}=0,\label{eq:vectorized-eq}
\end{equation}
where
\begin{equation*}
\bar{Q}=\left(\begin{matrix}
\rho \\
\rho u \\
\rho v \\
\rho w \\
e
\end{matrix}\right),
\;
\bar{E}=\left(\begin{matrix}
\rho u \\
\rho u^2+p \\
\rho uv \\
\rho uw \\
(e+p)u
\end{matrix}\right),
\;
\bar{F}=\left(\begin{matrix}
\rho v \\
\rho uv \\
\rho v^2+p \\
\rho vw \\
(e+p)v
\end{matrix}\right),
\end{equation*}
\begin{equation}
\bar{G}=\left(\begin{matrix}
\rho w \\
\rho uw \\
\rho vw \\
\rho w^2+p \\
(e+p)w
\end{matrix}\right),
\;
\bar{H}=\left(\begin{matrix}
0 \\
\rho \kappa_x \\
\rho \kappa_y \\
\rho \kappa_z \\
\rho(u\kappa_x+v\kappa_y+w\kappa_z)
\end{matrix}\right),
\end{equation}
in which $\rho$, $p$, $e$, $u$, $v$, $w$, $\kappa_x$, $\kappa_y$  and $\kappa_z$ are the density, pressure, total energy per unit volume, the $x$, $y$ and $z$ components of velocity and the $x$, $y$ and $z$ components of force, respectively. $p$ and $e$ are related as
\begin{equation}
p=(\gamma-1)\left[e-\frac{\rho}{2}(u^2+v^2+w^2)\right],
\end{equation}
in which $\gamma$ is the ratio of specific heats, since we only consider the adiabatic polytropic processes. The components of force are
\begin{eqnarray}
\kappa_x &=& -2v+\frac{\partial\psi}{\partial x},\label{eq:kappax}\\
\kappa_y &=& 2u+\frac{\partial\psi}{\partial y},\label{eq:kappay}\\
\kappa_z &=& \frac{\partial\psi}{\partial z},\label{eq:kappaz}
\end{eqnarray}
where $\psi$ is the dimensionless Roche effective potential, and Coriolis force has been combined into equations~(\ref{eq:kappax}) and (\ref{eq:kappay}) (The first terms on the right hand side of these equations). The formula of $\psi$ is
\begin{equation}
\psi=-\frac{a}{r_{\rm{WD}}}-\frac{qa}{r_{\rm{MS}}}-\frac{1+q}{2a^2}\left[\left(x-\frac{qa}{1+q}\right)^2+y^2\right],\label{eq:RochePot}
\end{equation}
where $q\,(=M_{\rm{MS}}/M_{\rm{WD}})$ is the mass ratio of the companion star to the WD (we focus on the situation where the companion star is a main-sequence star; see  Section~\ref{subsec:progenitor}), $a$ the distance between them, $r_{\rm{WD}}$ ($=\sqrt{x^2+y^2+z^2}$) and $r_{\rm{MS}}$ ($=\sqrt{(x-a)^2+y^2+z^2}$) the distances from a certain field point $(x,y,z)$ to the WD and the companion star, respectively.

\begin{figure}
	\includegraphics[width=\columnwidth]{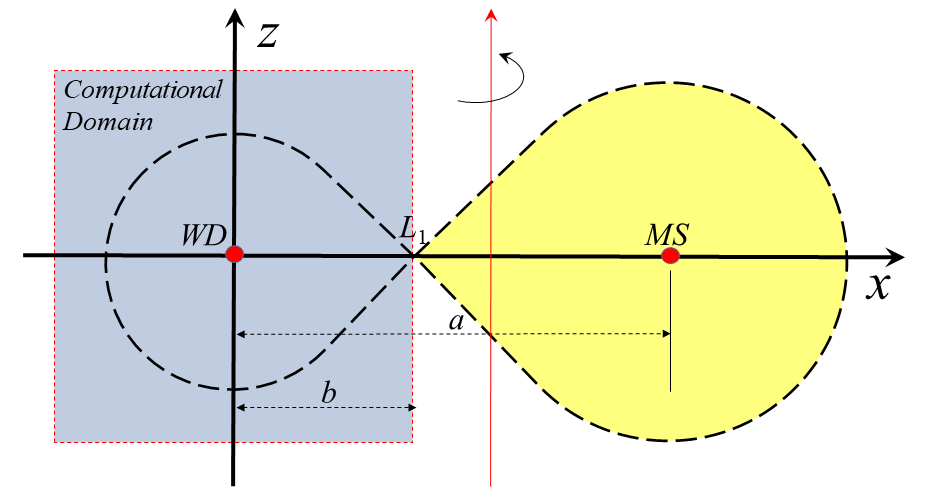}
	\caption{Geometry of the system.}
	\label{fig:Geometry}
\end{figure}

To investigate the detailed structure of the accretion flow, the simulations are focused on the region around the WD, in which the whole WD's partition of Roche lobe is contained (Figure~\ref{fig:Geometry}). We follow \cite{Makita2000} to assume that the gas is injected at the inner Lagrange point ($L_1$) between the WD and the companion. We choose twice the distance from $L_1$ to the WD centre, $2b$, as the length-scale, $\sqrt{G M_{\rm{WD}}/a}$ as the velocity-scale, and the density at $L_1$ as the density-scale, where $G$ is the gravitational constant. Under such scaling, the scaled $b$ has value $0.5$, and the Roche effective potential can be reduced to a simple formula (see equation (\ref{eq:RochePot})).

One thing worthy of note is that, while absolute values (e.g. in cgs units) of mass-related variables, such as density and mass outflow rate, depends on the mass transfer rate from the companion star, the scaled values in the simulation and the structure of the accretion flow do not. This is because under the adiabatic assumption, the accreted material is heated only by compression and work of gravitational force, which is proportional to density, so how the temperature changes only depends on the ratio the accreted gas is compressed and the distance it is accreted, not the actual amount of the accreted gas. If we consider a constant ratio of gas pressure to magnetic pressure (which is combined into $\gamma$ ignoring radiation pressure, see Section~\ref{subsec:gas} for details), then the total pressure is also proportional to density. As the force exerted on accreted material is a combination of pressure gradient and gravitational force, now we can see that the acceleration is irrelevant to actual density. As a result, the velocity profile is also irrelevant to actual density. This feature can be clearly seen in scaled equations, as even when the density of mass inflow at $L_1$ changes, the scaled equations remain the same.

To start the simulations, we need to know values of the mass of the WD $M_{\rm{WD}}$, the mass of the companion star $M_{\rm{MS}}$, the orbital period $P_{\rm{orb}}$, the ratio of specific heats $\gamma$, and temperature and velocity at $L_1$ point $T_{\rm{in}}$ and $V_{\rm{in}}$. $M_{\rm{WD}}$, $M_{\rm{MS}}$, $P_{\rm{orb}}$,  $T_{\rm{in}}$ and $V_{\rm{in}}$ are determined roughly according to previous theoretical and observational studies (see Section~\ref{subsec:progenitor} and~\ref{subsec:gas}). While their chosen values are only approximate, the simulation results are not sensible to their small changes based on our tests. The ratio of specific heats $\gamma$ (which eventually depends on the strength of magnetic field, see Section~\ref{subsec:gas}), on the other hand, has a large impact on the structure of the simulated accretion flow, especially the outflow, so we vary its value and discuss the simulation results in the paper.

The density at $L_1$ point, $\rho_{\rm{in}}$, does not influence the scaled results of the simulation. Thus we can calculate the mass outflow ratio
\begin{equation}\label{outflowratio}
\eta_{\rm{o}}=\frac{\dot{M}_{\rm{o}}}{\dot{M}_{\rm{t}}},
\end{equation}
where $\dot{M}_{\rm{o}}$ is the mass outflow rate and $\dot{M}_{\rm{t}}$ is the mass transfer rate from the companion star. Once we know the value of $\dot{M}_{\rm{o}}$ by other means (e.g., \cite{Fang2018} requires a mass outflow rate of $\sim10^{-7}M_{\odot}\rm{yr}^{-1}$), then $\dot{M}_{\rm{t}}$ and in turn $\rho_{\rm{in}}$ can be determined and absolute values of other mass-related variables can be obtained.

\subsection{The progenitor system}\label{subsec:progenitor}
As mentioned in the introduction, Tycho-G is proposed to be the surviving companion of the WD in Tycho's SN by multiple studies. Observations suggest that Tycho-G should be a type G0-G2 subgiant with mass $\sim1M_{\odot}$ \citep{Ruiz-Lapuente2004a}. Therefore, the surviving companion could have been a main-sequence star or a subgiant with mass larger than $1M_{\odot}$, because a surviving main-sequence star might resemble a subgiant but a subgiant would remain a subgiant after the explosion of SN Ia \citep{Marietta2000, Podsiadlowski2003, Ruiz-Lapuente2004a}. Since the lack of studies on the subgiant companion in the progenitor of SN, we only consider the main-sequence star (MS) as the companion in the progenitor system in our work \citep[therefore we only consider the WD+MS channel for SN Ia, see][for a review]{Wang2018}.

Researches in literature suggest that the minimum initial mass of the WD for producing SNe Ia in the WD+MS channel should be $\sim0.6M_{\odot}$ \citep{Han2004, Wang2010, Wang2018}. We choose a medium value $M_{\rm{WD}}=1M_{\odot}$ in our simulations to consider the intermediate state of a growing WD, which would be less extreme and better to reveal the average behavior of the progenitor evolution.

Combining the suggestions on the mass loss rate and the blowing duration from \cite{Zhou2016} ($\sim10^{-6}M_{\odot}\rm{yr}^{-1}$, $\sim4\times10^5\rm{yr}$) and \cite{Fang2018} ($\sim10^{-7}M_{\odot}\rm{yr}^{-1}$, $\sim10^5\rm{yr}$), the total mass carried out by the outflow during accretion should be in the range of $\sim10^{-2}M_{\odot}$ to $\sim0.4M_{\odot}$ before explosion. Considering the final WD mass of $\sim1.378M_{\odot}$ and a medium wind loss of $\sim0.2M_{\odot}$, we estimate the initial mass of the MS companion to be $M_{\rm{MS}}=1.6M_{\odot}$, which should be able to provide sufficient material for the progenitor to evolve into SN Ia and leave a surviving companion of $\sim1M_{\odot}$.

After $M_{\rm{WD}}$ and $M_{\rm{MS}}$, the third parameter, orbital period is determined as $\log(P_{\rm{orb}}/\rm{day})=-0.1$ ($P_{\rm{orb}}=0.794\,\rm{day}$) under the assumption that the Roche lobe is filled by the MS companion itself. It is worthwhile to point out that this triple of parameters, (1, 1.6, -0.1), represents a typical system in WD+MS channel, which was discussed by \cite{Wang2010} as an example for progenitor of SN Ia. Here, we use it again for the progenitor of Tycho's SN to reveal what kind of outflow can be blown out during accretion.

\subsection{The accreted material and magnetic field}\label{subsec:gas}
The material accreted by the WD comes from the surface of the type G companion, and thus should be a mixture of cool photospheric gas ($6000\sim7000\rm{K}$) and hot coronal gas ($>10^4 \rm{K}$). So it is cool, rarefied and might carry the frozen-in tangled magnetic field obtained from the surface of the companion star. In this kind of gas, the radiation pressure and cooling can be ignored due to its low temperature and low density. Therefore, we consider only the adiabatic polytropic processes in the accretion flow, and the corresponding polytropic exponential (ratio of specific heats) of gas-magnetic mixture is given by \citep{NY95}
\begin{equation}
\gamma=\frac{32-24\beta-3\beta^2}{24-21\beta},
\end{equation}
where $\beta\,(=p_{\rm{gas}}/p_{\rm{tot}})$ is the ratio of gas pressure to total pressure ($p_{\rm{tot}}=p_{\rm{gas}}+p_{\rm{mag}}$). Its value is in the range of 0 to 1, and the magnetic pressure becomes dominated when $\beta$ approaches 0 ($\gamma\rightarrow 4/3$), while gas pressure becomes dominated when $\beta$ approaches 1 \citep[$\gamma\rightarrow 5/3$, see][]{NY95}. The temperature of injected gas, $T_{\rm{in}}$, is assumed to be $10^4 \rm{K}$ (a mixture of cool photospheric gas and hot coronal gas), and its inflowing velocity is set to the local sound speed.

Note that the adiabatic assumption might be invalid in the inner region of the WD's accretion flow, where the temperature becomes higher, and $\beta$ might be non-constant in real gas due to magnetic dissipation. Therefore, our work in this paper is a simplified model, with which we try to investigate the overall picture of the accretion flow in the progenitor of Tycho's SN, and more detailed model will be considered with radiative MHD in our future work.

\section{Numerical Method}\label{Sec:Method}
All simulations in this paper are performed with the code Athena 4.2, which is in higher order Godunov method with finite volume discretization \citep{Stone2008}. Although there has been an updated Athena++, we choose the older, more mature Athena 4.2, which has such extensive documentation that it is easier for us to modify the code for our research than the new version. However, there are still some defects in the code that need further improvement in order to achieve our research goal in this paper. Below we discuss the settings we used to run our simulations and the modifications we made to Athena 4.2.
\subsection{Initial and boundary conditions}
The computational domain is a cubic box bounded with $-0.5\leqslant x, y, z\leqslant 0.5$, and the WD is put at origin, $L_1$ at $(0.5, 0, 0)$, and the centre of the companion star at $(a/2b, 0, 0)$ outside the box (based on our scaling, see Figure~\ref{fig:Geometry}.).

The computational domain has been evenly divided into $201\times 201\times 201$ cubic cells with size $2b/201=1.55\times 10^9\rm{cm}$, which is slightly larger than the diameter of the WD. All of these cells are initially filled by tenuous gas with zero-velocities, floor density $\rho_0=10^{-6}$ and floor energy $e_0=10^{-5}$ to mimic the interstellar medium. The origin of coordinates as well as the WD are placed in the central cell of the computational domain, which is treated as a empty hole through the function \textit{Userworkinloop} (named in Athena) to keep it at floor density and energy. The six square surfaces wrapping this hole is treated as the unidirectional inner boundaries (only flow into the hole, not any flow out of it, see also Section~\ref{subsec:UnidirBdry}).

In order to inject material into the computational box, we follow the treatment of \cite{Makita2000} to place a rectangular inlet at $L_1$ point. The injecting flow is assumed along the negative direction of $x$-axis with velocity about the sound speed \citep{Frank2002},
\begin{equation}
c_{\rm{s}}=\sqrt{\frac{\gamma \mathcal{R}T_{\rm{in}}}{\mu\beta}},
\end{equation}
where $\mathcal{R}$ is the ideal gas constant and $\mu$ the mean molecular weight, which is set to be 0.5 (pure ionized hydrogen).

The outer boundary is set on the six surfaces wrapping the whole computational domain (the surfaces with $x=\pm0.5$, $y=\pm0.5$ and $z=\pm0.5$, respectively). Except for the $L_1$ point, all of these surfaces are set with unidirectional outflow conditions, which can avoid any spurious effect from outside of the computational domain (see Section~\ref{subsec:UnidirBdry}).

\subsection{Unidirectional outflow boundary}\label{subsec:UnidirBdry}
The optional build-in outflow boundary condition in Athena 4.2 is the classical and simplest approach that it set all values of variables in ghost zones equal to the values in the corresponding active zones. This approach is exact for supersonic outflows but spurious reflection of waves will occur for subsonic outflows \citep{Stone1992}, as the inflow velocity would be copied to ghost zones even when the gas in active zones near the boundary is deflected inflow, which forms the actual inflow boundary and causes spurious mass inflow from those ghost zones. Thus we do not use the build-in outflow boundary condition but try to setup new functions to achieve a unidirectional outflow boundary.

In order to remove spurious reflection of waves, we make use of the entry of the problem-specific boundary condition in Athena 4.2 to setup our own functions to keep all ghost zones in the interstellar medium state defined by the density and internal energy with floor values, and three zero-velocities (they are not stationary but corotating with the system).

In order to avoid spurious mass inflow from ghost zones, we directly insert a function into the integrator loop to check the direction of the mass flux on the interface between each boundary cell and corresponding ghost cell. If spurious mass inflow is detected, we just remove all the numerical fluxes on this surface, which means variables of the mass, momentum and energy fluxes are reset to zero before they are used to update the cell-centred conservative variables.

Thus an unidirectional outflow boundary is established and we use it on the outer and inner boundaries. The treatment may deviate from astrophysical reality, but it can prevent any spurious reflection and inflow from influencing the computational domain with appropriate consumption of resources. Otherwise, one would need to setup a much larger computational domain to avoid the same problem.

\subsection{Recording exact fluxes}
In order to achieve our research goal in this paper, we need to obtain exact fluxes of mass, momentum and energy across the boundaries. There are two ways to do so, which are either post-processing of output data or interception and recording of numerical fluxes during calculation. We choose the latter because it is too difficult to exactly reconstruct the boundary fluxes, which are computed in Athena under some complicated physical settings (e.g. gravity, Coriolis' force, etc.).

In fact, the most exact fluxes are the instant numerical fluxes computed and used to update the dependent variables in Athena. Thus we insert a function into the loop of integration to intercept and record the instant numerical fluxes across the boundaries, which we technically define a new type of data file to store.

\subsection{Adding Coriolis' force}
We choose the Cartesian coordinates rather than the cylindrical ones, to avoid the polar singularity of curvilinear coordinates, which would affect the computational covering in the region around the polar axis. However, the build-in version of Coriolis' force in Athena 4.2 is only valid for cylindrical coordinates, so we need to add our own version of Coriolis' force in the calculation.

As shown by the basic equations (\ref{eq:vectorized-eq}) - (\ref{eq:kappaz}), Coriolis' force can not be treated as part of the effective potential like the gravitational and centrifugal forces, which can be added for the entry into stationary gravity in Athena (simply define a function in problem-generator of Athena to implement the Roche effective potential, equation (\ref{eq:RochePot})). Instead, we define a new source term according to equations (\ref{eq:kappax}) and (\ref{eq:kappay}) to implement Coriolis' force for Cartesian coordinates in Athena, and the build-in version is turned off in our calculation.

\section{Numerical results}\label{Sec:Result}

\begin{figure}
	\includegraphics[width=\columnwidth]{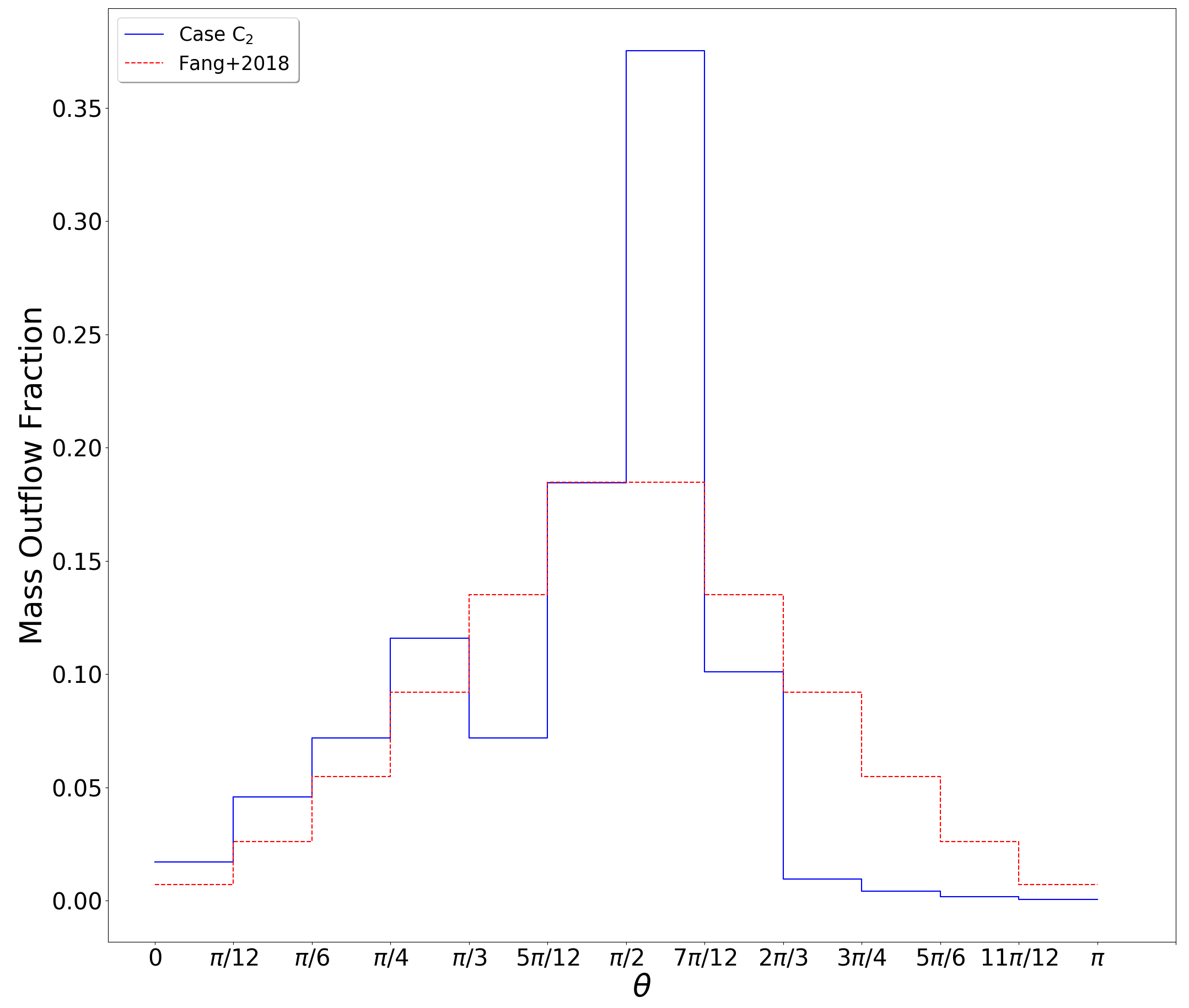}
	\caption{Fractions of total mass outflow of case $\rm{C_2}$. The dashed line is for the model used in {\protect\cite{Fang2018}}.}
	\label{fig:StairCaseC2}
\end{figure}

Following \cite{Zhou2016}, \cite{Fang2018} proposed that the supernova ejecta of Tycho's SN had evolved in a cavity blown by a latitude-dependent wind to reproduce the peculiar periphery of its remnant. The profile of this anisotropic wind was originally proposed in \cite{Rest2008} to model the tail of star Mira and it was also used to simulate the wind-blown cavity around the Kepler's SNR \citep{Toledo-Roy2014} and the Cygnus loop \citep{Fang2017}. In \cite{Fang2018}, they gave the profiles of wind density and velocity (Equations (1)-(4) in \cite{Fang2018}), which can be expressed simply as $\rho_{\rm{w}}\propto f(\theta)$ and $v_{\rm{w}}\propto [f(\theta)]^{-1/2}$ respectively, where $\theta$ is the polar angle, and
\begin{equation}
  f(\theta)=\xi-\left(\xi-1\right)\left|\cos\theta\right|^{1/2}.
\end{equation}
To facilitate comparison with our numerical results, we define the integral of mass outflow flux in a latitude band as
\begin{equation}\label{Eq:integmflux}
  I=2\pi r^2 \int_{\theta_l}^{\theta_u} \rho_{\rm{w}} v_{\rm{w}} \sin\theta d\theta \propto \int_{\theta_l}^{\theta_u} \sqrt{f(\theta)} \sin\theta d\theta\equiv F(\theta_l,\theta_u),
\end{equation}
where $r$ is the spherical radius, $\theta_l$ and $\theta_u$ are the lower and upper bounds of $\theta$ respectively, and the function $F(\theta_l,\theta_u)$ is the latitude profile of $I$. We set $\xi=20$ following \cite{Fang2018} to reproduce their profile and divide the outer boundary of computational domain into 12 latitude bands with width $\Delta\theta=\pi/12$. In this way, the latitude profile implemented in \cite{Fang2018} can be visualized as a step function and used in comparison with our results (see Figures~\ref{fig:StairCaseA},~\ref{fig:StairCaseB1},~\ref{fig:StairCaseB2},~\ref{fig:StairCaseC1} \&~\ref{fig:StairCaseC2}).

In this paper, we show simulation results for five different cases (Table \ref{tab:DiffCases}) of the progenitor model discussed in Section \ref{Sec:Model} to investigate the latitude-dependent outflow. In each case, the outflow ratio $\eta_{\rm{o}}$ is calculated. \cite{Fang2018} requires a total mass outflow rate of $10^{-7}M_{\odot}\rm{yr}^{-1}$ for a duration of $\sim10^5\rm{yr}$, so we can calculate the mass transfer rate from the companion star and the strength of magnetic field in the accreted material for \cite{Fang2018} in each case. For comparison, we also calculate the mass transfer rate and magnetic field strength for \cite{Zhou2016} (which proposes a total mass outflow rate of $\sim10^{-6}M_{\odot}\rm{yr}^{-1}$ for a duration of $\sim4\times10^5\rm{yr}$).

\begin{table}
	\caption{Parameters for the simulations. The common parameters are $M_{\rm{WD}}=1M_{\odot}$, $M_{\rm{MS}}=1.6M_{\odot}$, $P_{\rm{orb}}=0.794\rm{day}$ and $T_{\rm{in}}=10^4\rm{K}$.}
	\label{tab:DiffCases}
	\begin{tabular}{llll}
		\hline
		Models & $\beta$ & $\gamma$ \\
		\hline
		A & 1 & 1.6667 (5/3)\\
		$\rm{B_1}$ & 0.5 & 1.4259\\
		$\rm{B_2}$ & 0.5 & 1.426\\
		$\rm{C_1}$ & 0.72 & 1.482\\
		$\rm{C_2}$ & 0.72 & 1.481\\
		\hline
	\end{tabular}	
\end{table}

\subsection{Case A}

\begin{figure}
	\includegraphics[width=\columnwidth]{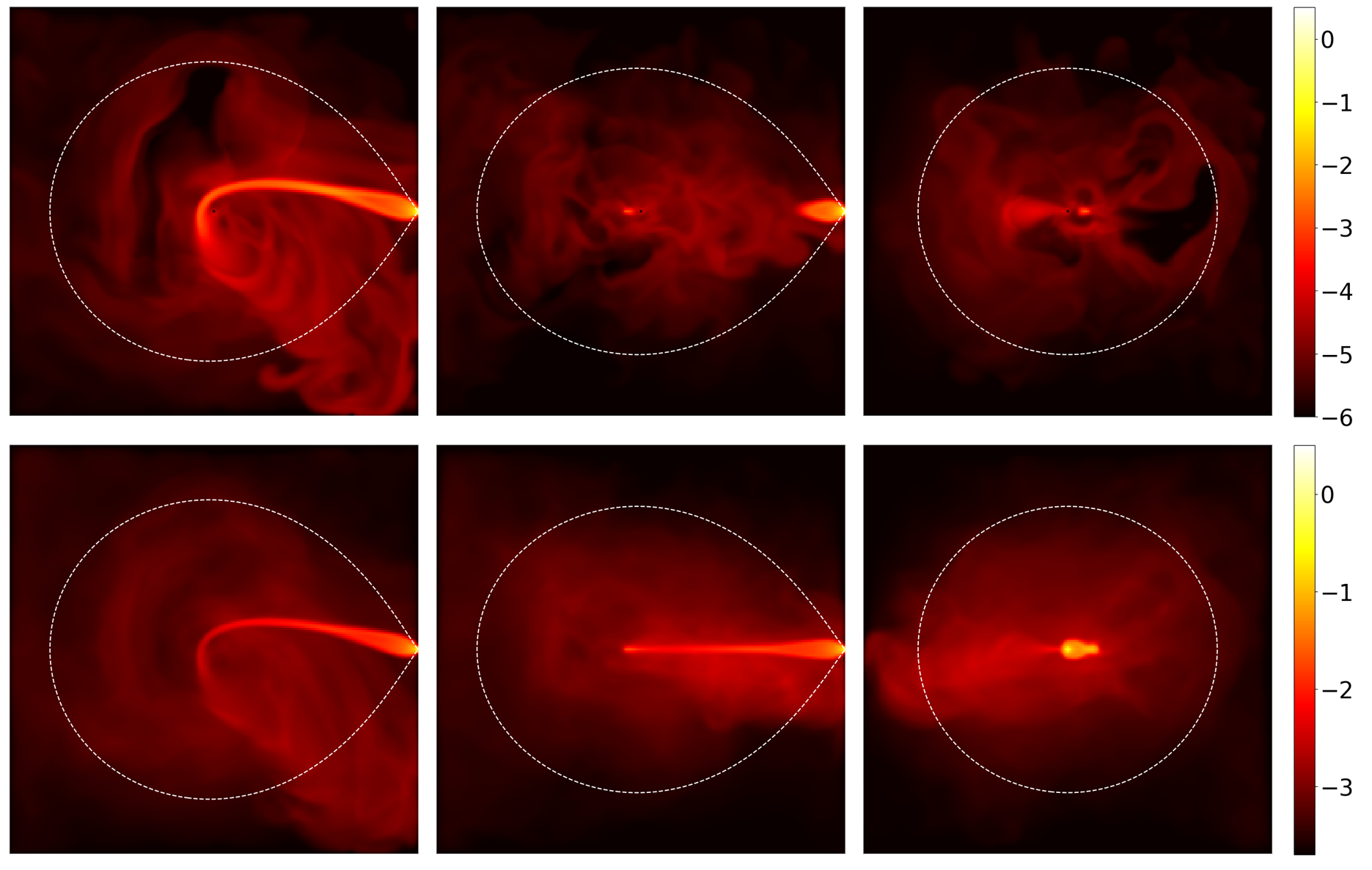}
	\caption{Images of density at the final time of case A. The three upper subplots are the central slices extracted from the cardinal planes with $z=0$, $y=0$, and $x=0$ from left to right respectively. The three lower subplots are the corresponding integral images along the same line of sight. The white dashed line is the projection of the Roche lobe on the plane perpendicular to the line of sight. The density and column density in the subplots are calibrated with logarithms of scaled values.}
	\label{fig:SlicesCaseA}
\end{figure}

\begin{figure}
	\includegraphics[width=\columnwidth]{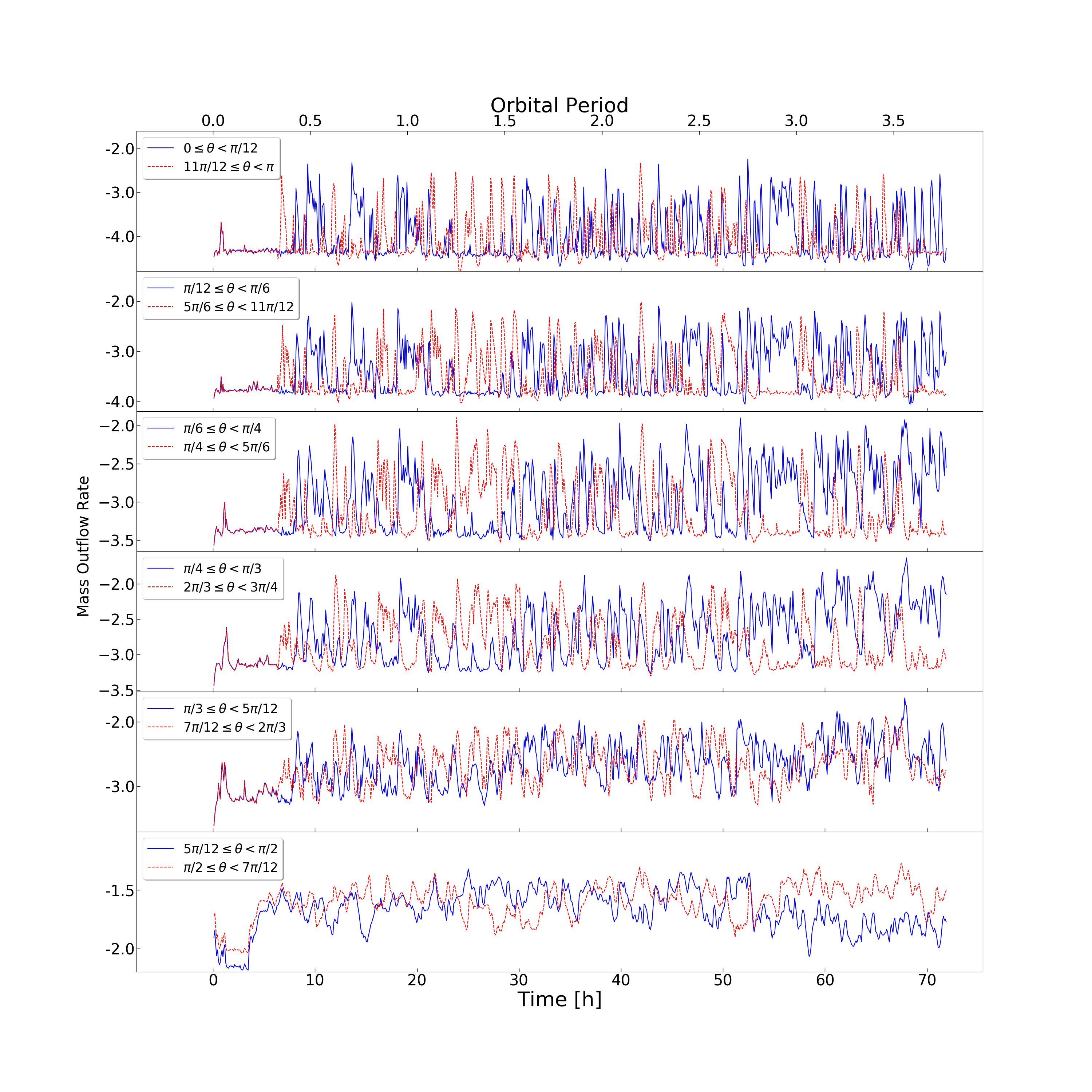}
	\caption{Variation of latitude-dependent mass outflow in case A. We divide the entire outer boundary into 12 latitude-bands by the polar angle $\theta$. Each subplot contains the curves from two symmetric bands and the subplots are arranged from top to bottom in order from the pole to the equator. The ordinate indicates logarithms of scaled mass outflow rate.}
	\label{fig:MassOutflowCaseA}
\end{figure}

\begin{figure}
	\includegraphics[width=\columnwidth]{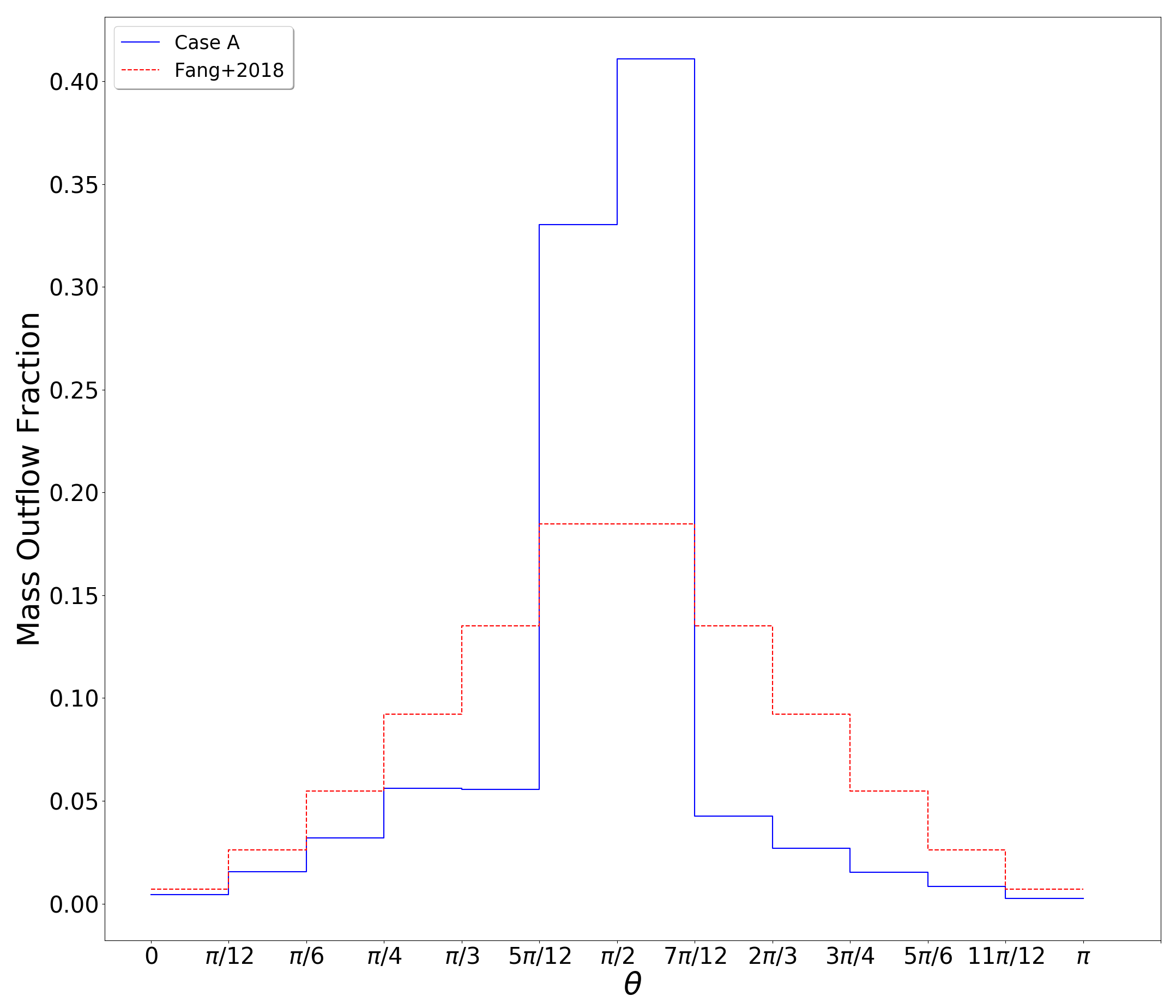}
	\caption{Fractions of total mass outflow from the latitude bands of case A, during the quasi-steady state of accretion. The dashed line is for the outflow model used in {\protect\cite{Fang2018}}.}
	\label{fig:StairCaseA}
\end{figure}

Case A with $\beta=1$ is when the accreted material has no frozen-in tangled magnetic field or the intensity of field is negligible. This simulation has been run for over 70 hours of observer time ($>3.5P_{\rm{orb}}$), which is long enough to achieve quasi-steady accretion. We show the images of density at final time in Figure~\ref{fig:SlicesCaseA}, which are composed of three central slices (three upper subplots) and the integral images along the corresponding line of sight (three lower subplots). Figure~\ref{fig:MassOutflowCaseA} shows the variation of mass outflow rate in 12 latitude bands, where the curves from each pair of symmetric bands are drawn in the same subplot. We can see that a strong outflow is generated near the equatorial plane, and while outflows also exist in other directions, they are extremely weak compared with the equatorial one. This can be clearly seen in Figure~\ref{fig:StairCaseA}, which shows the relevant fractions of total mass outflow from the 12 bands during the quasi-steady state, obtained after 30 hours of observer time. The equatorial 
outflow disrupts the formation of a compact accretion disk, so no apparent disk can be seen in Figure~\ref{fig:SlicesCaseA}.

Compared with the latitude-dependent wind of \cite{Fang2018}, the outflow fractions of two equatorial bands in case A are much larger than their counterparts (dashed line in Figure~\ref{fig:StairCaseA}), and the fractions of other bands except the two polar bands are much lower than those in \cite{Fang2018}. Therefore, case A is not the case required for the latitude-dependent wind of \cite{Fang2018}. 

The mean mass outflow ratio during quasi-steady accretion in case A is 96.4\%, which means that 96.4\% of the mass transferred from the MS companion is lost during accretion in the form of outflow. Disregarding the difference in latitude profile of the outflow, case A would require a mass transfer rate of $1.04\times10^{-7}M_{\odot}\rm{yr}^{-1}$ to blow away the same amount of material as the wind of \cite{Fang2018}, and 	$1.04\times10^{-6}M_{\odot}\rm{yr}^{-1}$ as that of \cite{Zhou2016}. While \cite{Zhou2016} is an observational paper and lays the foundation of numerical simulation research of \cite{Fang2018}, their estimation of mass outflow rate is based on the model of \cite{Han2004}, in which there is no explicit consideration on the anisotropy of outflow (i.e. the outflow is spherically symmetric), which would require a larger outflow rate than the anisotropic one of \cite{Fang2018}.

\subsection{Case B}\label{caseB}
\begin{figure}
	\includegraphics[width=\columnwidth]{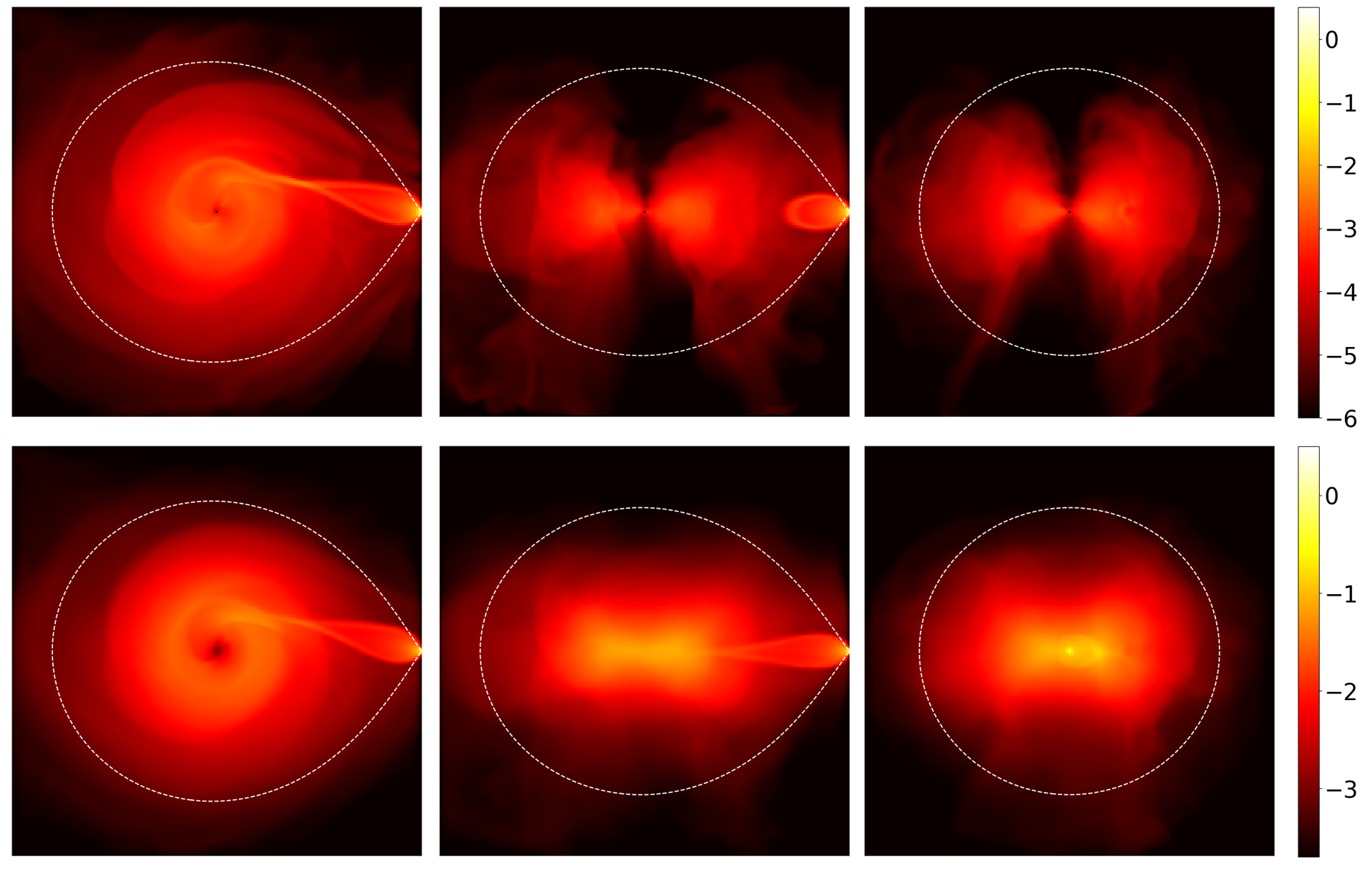}
	\caption{Images of density at the final time of case $\rm{B_1}$. The arrangement of subplots is the same as Figure~\ref{fig:SlicesCaseA}.}
	\label{fig:SlicesCaseB1}
\end{figure}

\begin{figure}
	\includegraphics[width=\columnwidth]{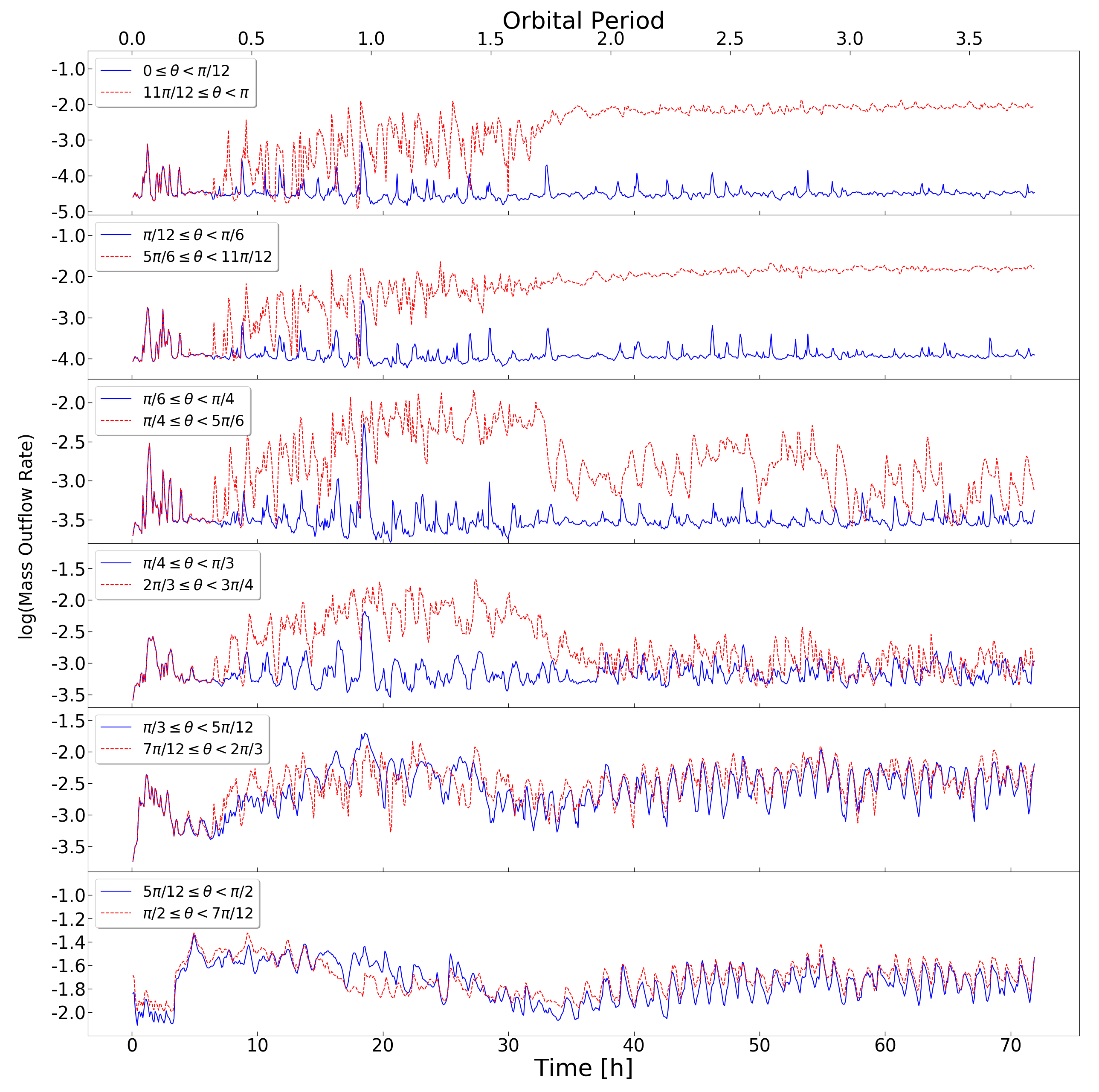}
	\caption{Variation of latitude-dependent mass outflow of case $\rm{B_1}$. The arrangement of subplots is the same as Figure~\ref{fig:MassOutflowCaseA}.}
	\label{fig:MassOutflowCaseB1}
\end{figure}

\begin{figure}
	\includegraphics[width=\columnwidth]{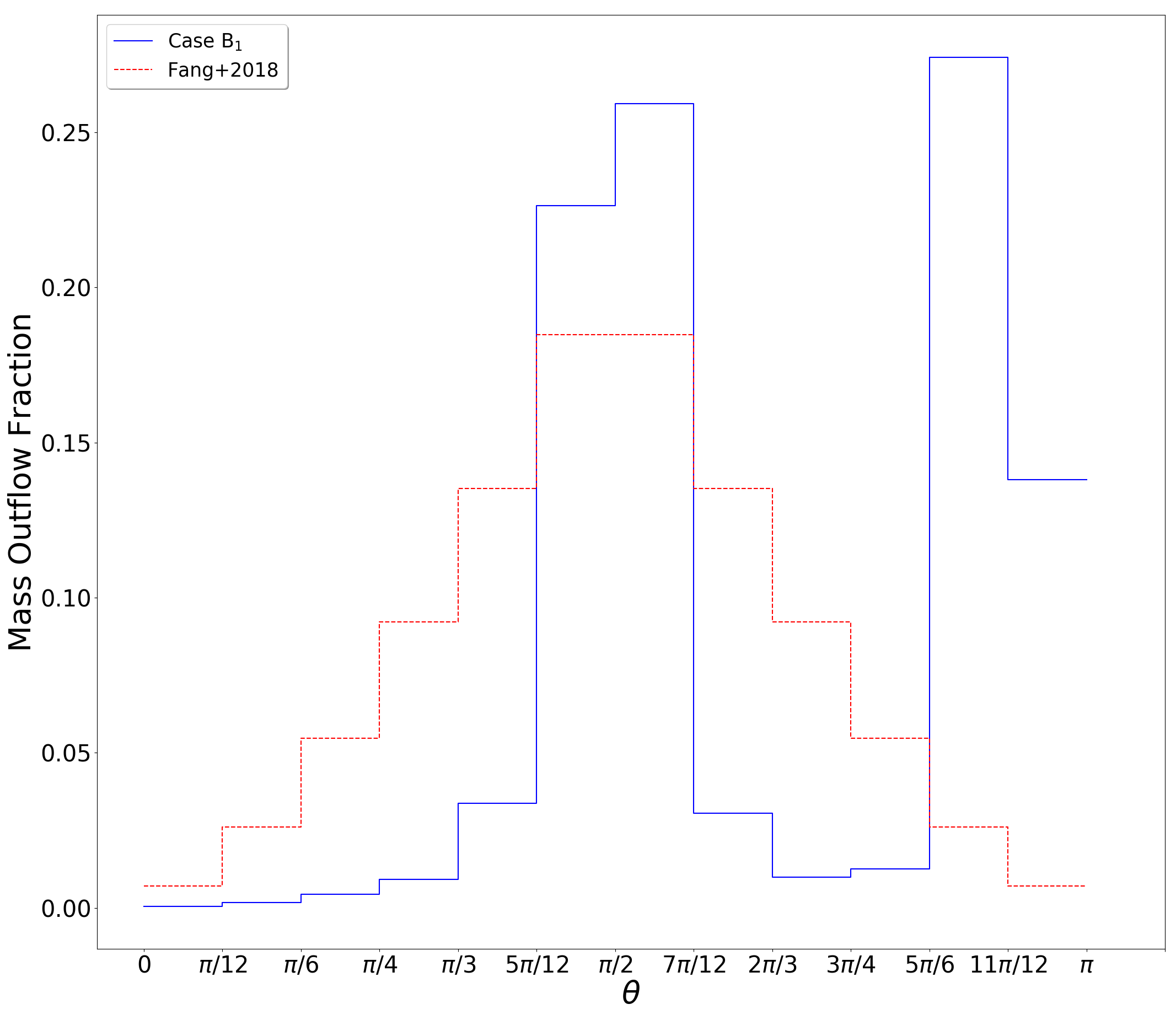}
	\caption{Fractions of total mass outflow from the latitude bands of case $\rm{B_1}$, during the quasi-steady state of accretion. The dashed line is for the model used in {\protect\cite{Fang2018}}.}
	\label{fig:StairCaseB1}
\end{figure}

Case $\rm{B_1}$ with $\beta=0.5$ is when the magnetic energy and internal energy in accreted material are equal (i.e. energy equipartition). Figure~\ref{fig:SlicesCaseB1} shows the the images of density at final time of the simulation run, when the accretion flow has reached quasi-steady state. Figures~\ref{fig:MassOutflowCaseB1} and \ref{fig:StairCaseB1} show the mass outflow rates and their fractions in the latitude bands of case $\rm{B_1}$, just like Figures~\ref{fig:MassOutflowCaseA} and \ref{fig:StairCaseA} for case A. A strong equatorial outflow can still be seen in these figures. However, the lower half of the computational domain has flow extending to the bottom boundary (Figure~\ref{fig:SlicesCaseB1}), which is another outflow in comparable strength with the equatorial one and peaks in the band $5\pi/6<\theta<11\pi/12$ (Figures~\ref{fig:MassOutflowCaseB1} and \ref{fig:StairCaseB1}). This polar outflow carries away part of the mass, energy and angular momentum and helps stablizing the accretion flow on the equator, and an apparent accretion disk is formed on the equator (Figure~\ref{fig:SlicesCaseB1}) as a result.

\begin{figure}
	\includegraphics[width=\columnwidth]{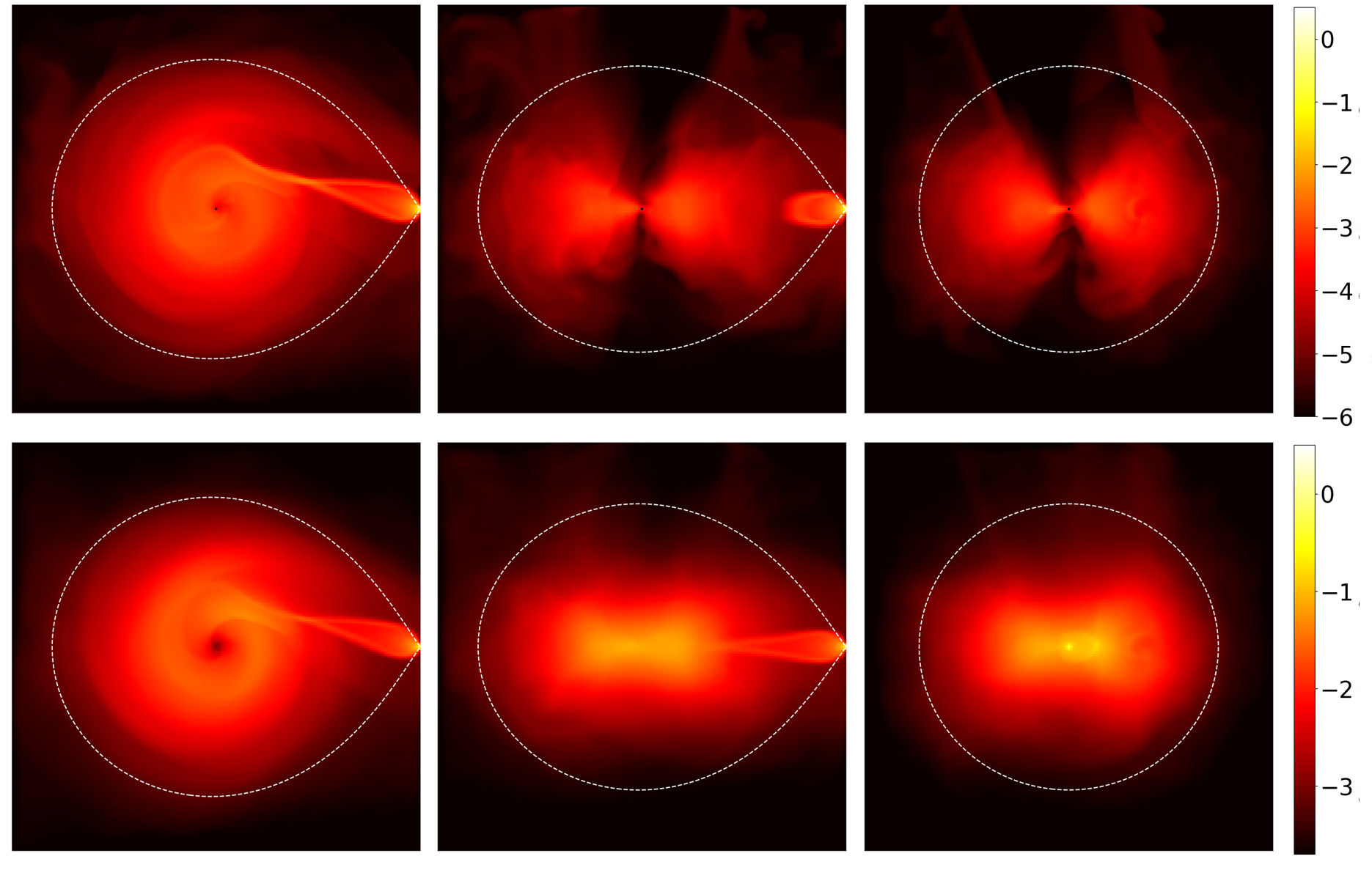}
	\caption{Images of density at the final time of case $\rm{B_2}$. The arrangement of subplots is the same as Figure~\ref{fig:SlicesCaseA}.}
	\label{fig:SlicesCaseB2}
\end{figure}

\begin{figure}
	\includegraphics[width=\columnwidth]{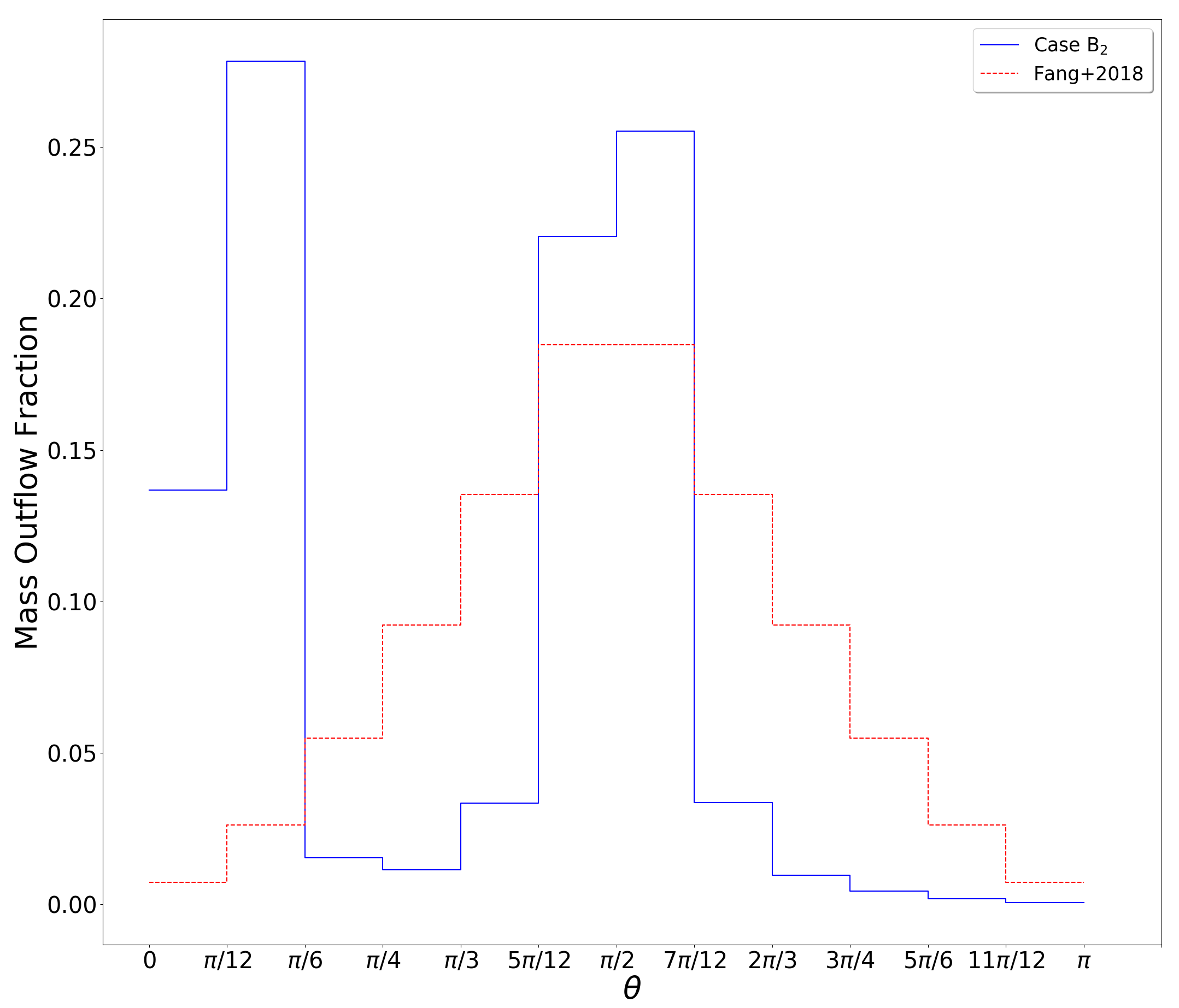}
	\caption{Fractions of total mass outflow from the latitude bands of case $\rm{B_2}$, during the quasi-steady state of accretion. The dashed line is for the model used in {\protect\cite{Fang2018}}.}
	\label{fig:StairCaseB2}
\end{figure}

The polar outflow in case $\rm{B_1}$ is asymmetric, as there is no such outflow in the upper half of the computational domain. It should be noted that the direction of the polar outflow is somewhat stochastic in the simulation, as shown in case $\rm{B_2}$, which has been performed with all the same parameters and settings as case $\rm{B_1}$ except the parameter $\gamma$ with a little difference in value (see Table~\ref{tab:DiffCases}). The results of this simulation (Figures~\ref{fig:SlicesCaseB2} and \ref{fig:StairCaseB2}) show a polar outflow with similar structure in the upper half of the computational domain instead. Based on these results, we suspect that such an asymmetric outflow feature is due to the chaotic effect caused by our assumption of a constant $\gamma$. In real astrophysical environment, the strength of magnetic field in accreted material is impossible to keep strictly constant everywhere as our model assumes. It means that the actual value of $\beta$ and the corresponding value of $\gamma$ will fluctuate, which will lead to a stochastic change in the outflow direction and eventually form an overall symmetric outflow structure. Therefore, we think that our asymmetric outflow can still be used in comparison with the symmetric one of \cite{Fang2018}, which is shown in Figures~\ref{fig:StairCaseB1} and \ref{fig:StairCaseB2}. We can see that the simulation results deviate a lot from the outflow structure in \cite{Fang2018}, which shows no sign of such a concentrated polar outflow. Therefore, the equipartition case is not the case required for the latitude-dependent wind of \cite{Fang2018} either. 

The mean mass outflow ratio during quasi-steady accretion in case $\rm{B_1}/\rm{B_2}$ is 90.0\%, so case $\rm{B_1}/\rm{B_2}$ would require a mass transfer rate of $1.11\times10^{-7}M_{\odot}\rm{yr}^{-1}$ to blow away the same amount of material as the wind of \cite{Fang2018}, and the strength of magnetic field in accreted material is calculated to be $8.16\times10^3\rm{G}$. For \cite{Zhou2016}, these two values are $1.11\times10^{-6}M_{\odot}\rm{yr}^{-1}$ and $2.58\times10^4\rm{G}$, respectively.

\subsection{Case C}
\begin{figure}
	\includegraphics[width=\columnwidth]{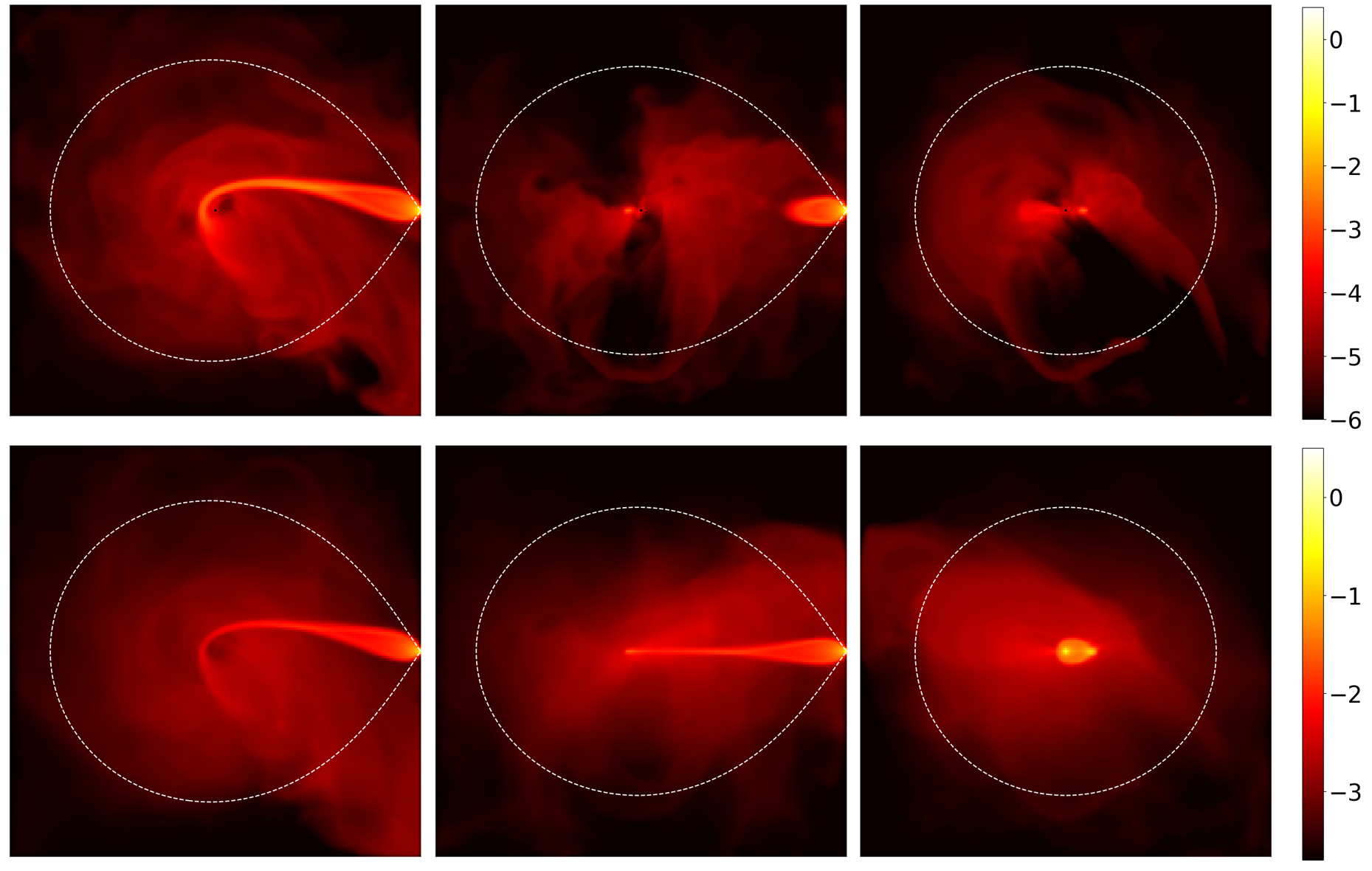}
	\caption{Images of density at the final time of case $\rm{C_1}$. The arrangement of subplots is the same as Figure~\ref{fig:SlicesCaseA}.}
	\label{fig:SlicesCaseC1}
\end{figure}

\begin{figure}
	\includegraphics[width=\columnwidth]{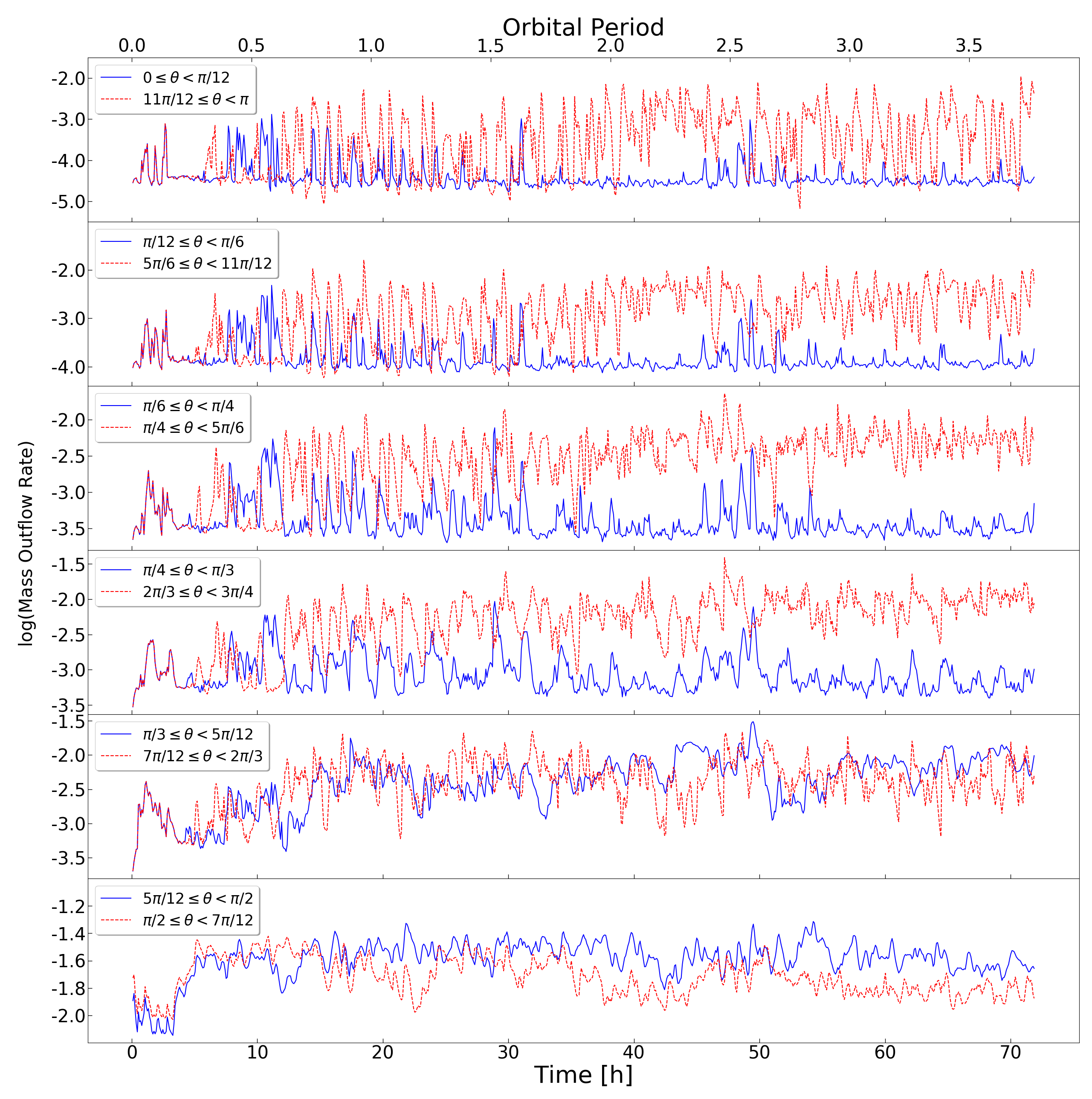}
	\caption{Variation of latitude-dependent mass outflow of case $\rm{C_1}$. The arrangement of subplots is the same as Figure~\ref{fig:MassOutflowCaseA}.}
	\label{fig:MassOutflowCaseC1}
\end{figure}

\begin{figure}
	\includegraphics[width=\columnwidth]{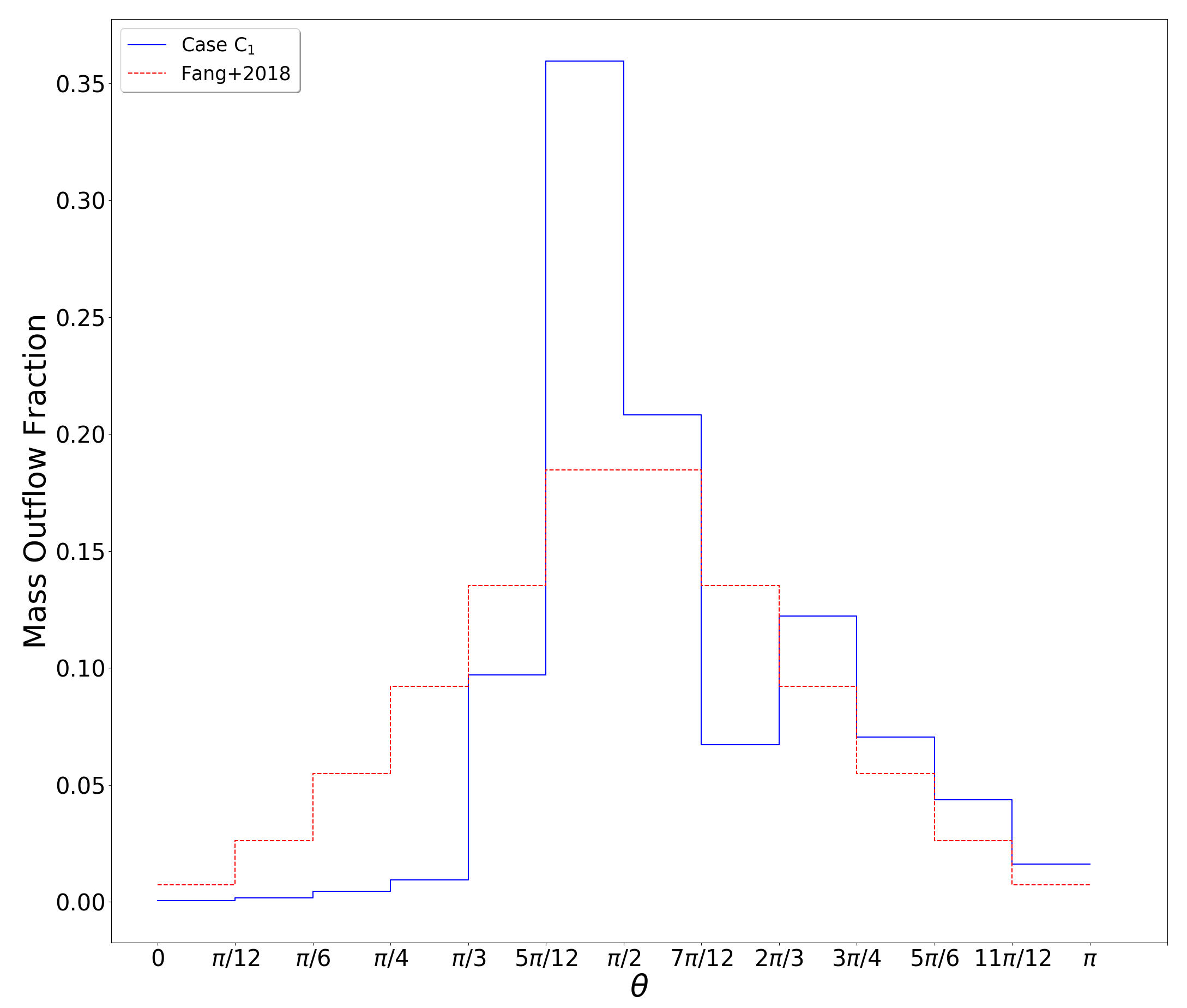}
	\caption{Fractions of total mass outflow of case $\rm{C_1}$. The dashed line is for the model used in {\protect\cite{Fang2018}}.}
	\label{fig:StairCaseC1}
\end{figure}

Case $\rm{C_1}$ is the case of which the outflow distribution is closest to that of \cite{Fang2018} in our simulation runs, obtained by testing various values of $\beta$ parameter with the method of bisection. It has $\beta=0.72$, so the strength of magnetic field lies between case A and case $\rm{B_1}/\rm{B_2}$. Just like in case A, no apparent disk is formed (Figure~\ref{fig:SlicesCaseC1}). However, it also has the feature of asymmetric outflow away from the equatorial outflow, which is similar to case $\rm{B_1}/\rm{B_2}$, though the outflow is not as concentrated, not peaked just around the pole, and ranges across a much larger polar angle (Figures~\ref{fig:MassOutflowCaseC1} and~\ref{fig:StairCaseC1}). The distribution of outflow in the lower half of the computational domain is roughly in agreement with that of \cite{Fang2018} (dashed line in Figure~\ref{fig:StairCaseC1}). As their distribution is a preset formula, it declines smoothly from the outflow maximum on the equator to the minimum around the pole. Our simulation, however, finds out that either the outflow is mostly concentrated around the equatorial plane when the magnetic field is negligible, or another major outflow is generated away from the equator when the magnetic field is strong enough. This seems more realistic than the formula of \cite{Fang2018}, and as the two distributions roughly agree, we suspect that if their calculation is repeated with our outflow structure in the lower half of the computational domain in case C1 as the input, the periphery of Tycho's SNR could also be roughly reproduced.

The mean mass outflow ratio during quasi-steady accretion in case $\rm{C_1}$ is 95.0\%. The mass transfer rate from the companion and the strength of magnetic field are $1.05\times10^{-7}M_{\odot}\rm{yr}^{-1}$ and $5.44\times10^3 \rm{G}$ for \cite{Fang2018}, respectively. We can see that the strength of magnetic field is of the same magnitude as those in sunspots \citep[e.g.,][]{Wang2019, Siu2019}. As the accreted material comes from a MS companion, this is self-consistent. For comparison, we also calculate these two values for \cite{Zhou2016}, which are $1.05\times10^{-6}M_{\odot}\rm{yr}^{-1}$ and $1.72\times10^4\rm{G}$, respectively.

Like case $\rm{B_1}$, a shadow simulation of case $\rm{C_1}$, case $\rm{C_2}$, is performed, in which all of the parameters are the same except $\gamma$ with a little bit difference. The result is generally a mirror of case $\rm{C_1}$. As mentioned in Section~\ref{caseB}, we suspect that it is due to our assumption of a constant $\gamma$, and in real situations the outflow structure would be symmetric.

\section{Summary and Discussion}\label{Sec:Summary}


We run 3D simulations of the accretion flow in the SD model (WD+MS channel) for Tycho' SN. The mass of the WD, mass of the companion star, and the orbital period are set to be 1$M_{\odot}$, 1.6$M_{\odot}$, and 0.794 day, respectively, according to previous theoretical and observational studies. The inflowing gas from the MS companion is set to be of $10^4$K with a velocity of the local sound speed, which are natural settings for the WD+MS channel. The gas pressure ratio $\beta$  $(=p_{\rm{gas}}/(p_{\rm{gas}}+p_{\rm{mag}}))$ in the accreted material, and consequently the strength of magnetic field under the adiabatic assumption, is varied and quasi-steady states of the accretion flow are obtained in multiple simulation runs. When the magnetic field in the accreted material is negligible, the simulation shows strong outflow near the equatorial plane, which disrupts the formation of a compact accretion disk, as shown in case A. Outflows in other directions are very weak compared with the equatorial one. As the magnetic field becomes stronger, another major outflow is generated at higher latitudes away from the equatorial one, which carries away part of the mass, energy and angular momentum and helps stablizing the accretion flow on the equator. Eventually this outflow will become strong enough and an apparent equatorial accretion disk will be formed, as shown in case $\rm{B_1}/\rm{B_2}$. In both case A and case $\rm{B_1}/\rm{B_2}$, the outflow distribution deviates much from the latitude-dependent wind required to form the peculiar periphery of Tycho's SNR \citep{Fang2018}, which means that in the progenitor of Tycho's SN the accreted material from the companion star must contain some magnetic field, yet the magnetic field cannot be too strong to be in energy equipartition with the internal energy ($B=8.16\times10^3\rm{G}$ in our simulation). The closest result to the latitude-dependent wind of \cite{Fang2018} in our simulation is obtained when $\beta$ is set to be 0.72, with a magnetic field $B=5.44\times10^3 \rm{G}$. This field strength is of the same magnitude as those in sunspots, and thus is plausible for the WD+MS channel.

Scaled equations are used in our calculation, which are independent of the absolute density of accreted material. The mass outflow ratio is calculated in each case, and absolute values of mass transfer rate from the companion star $\dot{M}_{\rm{t}}$ as well as other mass-related variables (e.g. the strength of magnetic field $B$) can be calculated once the mass outflow rate is provided. \cite{Fang2018} proposes a total mass outflow rate of $10^{-7}M_{\odot}\rm{yr}^{-1}$ for a duration of $\sim10^5\rm{yr}$, while \cite{Zhou2016} proposes a total mass outflow rate of $\sim10^{-6}M_{\odot}\rm{yr}^{-1}$ for a duration of $\sim4\times10^5\rm{yr}$, which is much larger. The reason is that while \cite{Zhou2016} provides observations of the expanding bubble driven by the progenitor of Tycho's SN, their estimation of the mass outflow rate is based on a spherically symmetric outflow model, which requires a larger outflow rate than the latitude-dependent outflow model of \cite{Fang2018}. Consequently the corresponding $\dot{M}_{\rm{t}}$ and $B$ calculated for \cite{Zhou2016} is considerably larger than those for \cite{Fang2018} in every case except for case A in which $B=0$.

The outflow considered in this paper provides another source of mass loss than the mass loss caused by hydrogen and helium flashes on the WD surface which are often considered in binary evolution researches \citep[see the review by][and references therein]{Wang2018}. The mass loss ratio $\eta_{\rm{o}}$ is extremely large (above 90\%), yet it is consistent with researches in accretion physics \citep[e.g.,][]{Begelman2012}, and this outflow only lasts for a limited time before the SN explosion, so it does not handicap the mass accumulation of the WD much (for \cite{Fang2018}, the outflow only takes away $\sim10^{-2}M_{\odot}$ altogether; even for \cite{Zhou2016} in which the outflow is much larger, the total mass taken away is $\sim0.4M_{\odot}$).

The asymmetric distribution of outflow between upper and lower halves of the computational domain in our simulation is probably due to assumption of a constant ratio of specific heats $\gamma$. It is a chaotic effect that a little difference in $\gamma$ value may finally change the direction of the polar outflow. We suspect that the distribution of outflow will become symmetric under the fluctuation of $\gamma$ (i.e. fluctuation of tangled magnetic field) in real situations. The stochastic fluctuation of tangled magnetic field should be considered in further researches with magneto-hydrodynamical simulation.

\section*{Acknowledgements}

We thank Bo Wang for helpful discussion. This work is supported by the Natural Science Foundation of China under grants No. 11373002 and 11703083, and Natural Science Foundation of Fujian Province of China under grant No. 2018J01007. The calculations were performed on TianHe-1(A) at National Supercomputer Center in Tianjin.









\bsp	
\label{lastpage}
\end{document}